\documentclass[aip,rsi,reprint]{revtex4-1}

\usepackage{color}
\usepackage{graphicx}
\usepackage{float}
\usepackage[normalem]{ulem}
\usepackage{hyperref}
\hypersetup{
    colorlinks=true,
    linkcolor=blue,
    filecolor=blue,      
    urlcolor=blue,
    citecolor=blue
}

\begin{document}

\title{Expanding the momentum field of view in angle-resolved photoemission systems with hemispherical analyzers}

\author{Nicolas Gauthier}
\thanks{The authors to whom correspondence may be addressed: \href{mailto:nicolas.gauthier5@usherbrooke.ca}{nicolas.gauthier5@usherbrooke.ca}, \href{mailto:kirchman@stanford.edu}{kirchman@stanford.edu}.}
\affiliation{Stanford Institute for Materials and Energy Sciences, SLAC National Accelerator Laboratory, Menlo Park, California 94025, USA}
\affiliation{Geballe Laboratory for Advanced Materials, Departments of Applied Physics, Stanford University, Stanford, California 94305, USA}
\affiliation{Departments of Physics, Stanford University, Stanford, California 94305, USA}
\author{Jonathan A. Sobota}
\affiliation{Stanford Institute for Materials and Energy Sciences, SLAC National Accelerator Laboratory, Menlo Park, California 94025, USA}
\author{Heike Pfau}
\affiliation{Stanford Institute for Materials and Energy Sciences, SLAC National Accelerator Laboratory, Menlo Park, California 94025, USA}
\affiliation{Advanced Light Source, Lawrence Berkeley National Laboratory, Berkeley, California 94720, USA}
\author{Alexandre Gauthier}
\affiliation{Stanford Institute for Materials and Energy Sciences, SLAC National Accelerator Laboratory, Menlo Park, California 94025, USA}
\affiliation{Geballe Laboratory for Advanced Materials, Departments of Applied Physics, Stanford University, Stanford, California 94305, USA}
\affiliation{Departments of Physics, Stanford University, Stanford, California 94305, USA}
\author{Hadas Soifer}
\affiliation{School of Physics and Astronomy, Tel-Aviv University, 69978, Israel}
\author{Maja~D.~Bachmann}
\affiliation{Geballe Laboratory for Advanced Materials, Departments of Applied Physics, Stanford University, Stanford, California 94305, USA}
\author{Ian R. Fisher}
\affiliation{Stanford Institute for Materials and Energy Sciences, SLAC National Accelerator Laboratory, Menlo Park, California 94025, USA}
\affiliation{Geballe Laboratory for Advanced Materials, Departments of Applied Physics, Stanford University, Stanford, California 94305, USA}
\author{Zhi-Xun Shen}
\affiliation{Stanford Institute for Materials and Energy Sciences, SLAC National Accelerator Laboratory, Menlo Park, California 94025, USA}
\affiliation{Geballe Laboratory for Advanced Materials, Departments of Applied Physics, Stanford University, Stanford, California 94305, USA}
\affiliation{Departments of Physics, Stanford University, Stanford, California 94305, USA}
\author{Patrick S. Kirchmann}
\thanks{The authors to whom correspondence may be addressed: \href{mailto:nicolas.gauthier5@usherbrooke.ca}{nicolas.gauthier5@usherbrooke.ca}, \href{mailto:kirchman@stanford.edu}{kirchman@stanford.edu}.}
\affiliation{Stanford Institute for Materials and Energy Sciences, SLAC National Accelerator Laboratory, Menlo Park, California 94025, USA}

\date{\today}

\begin{abstract}
In photoelectron spectroscopy, the measured electron momentum range is intrinsically related to the excitation photon energy. Low photon energies $<10$~eV are commonly encountered in laser-based photoemission and lead to a momentum range that is smaller than the Brillouin zones of most materials. This can become a limiting factor when studying condensed matter with laser-based photoemission. An additional restriction is introduced by widely used hemispherical analyzers that record only electrons photoemitted in a solid angle set by the aperture size at the analyzer entrance. Here, we present an upgrade to increase the effective solid angle that is measured with a hemispherical analyzer. We achieve this by accelerating the photoelectrons towards the analyzer with an electric field that is generated by a bias voltage on the sample. Our experimental geometry is comparable to a parallel plate capacitor and, therefore, we approximate the electric field to be uniform along the photoelectron trajectory. With this assumption, we developed an analytic, parameter-free model that relates the measured angles to the electron momenta in the solid and verify its validity by comparing with experimental results on the charge density wave material TbTe$_3$. By providing a larger field of view in momentum space, our approach using a bias potential considerably expands the flexibility of laser-based photoemission setups. 
\end{abstract}

\maketitle

%%%%%%%%%%%%%%
%%% INTRO %%%
%%%%%%%%%%%%%%

\section{Introduction}
\label{secIntro}

Angle-resolved photoemission spectroscopy (ARPES) probes the electronic structure in momentum space and plays a fundamental role in our understanding of material properties.\cite{Sobota2020} One important consideration when designing experiments is the accessible momentum range. 
In a conventional ARPES experiment, the in-plane momentum of the electrons in the solid $k_\parallel^S$ is determined by measuring the emission angle $\theta_S$ at the sample surface and the kinetic energy $E_K^S$ of the photoemitted electrons. Based on energy and momentum conservation, the in-plane momentum is then determined by\cite{Hue95,Suga2014}
\begin{equation}
k_\parallel^S = \frac{1}{\hbar} \sqrt{2m E_K^S} \sin \theta_S,
\label{eq1}
\end{equation}
where $m$ is the free electron mass. 

$E_K^S$ varies with the excitation photon energies, which can vary from the ultraviolet to the hard x-ray regime. Higher photon energies result in higher kinetic energies and therefore in larger accessible momenta. In contrast, low photon energies access a narrower momentum range but typically have better momentum resolution.\cite{Koralek07,Vishik2010,Xiong2017,Tamai2019}

The measured momentum range also depends on the experimental geometry and the photoelectron detectors. Conventional hemispherical imaging analyzers are widely used in ARPES setups to measure the photoelectron's emission angle and kinetic energy. The emission angle is measured along a single direction, as defined by the analyzer slit, and the angular range is geometrically limited by a fixed aperture at the entrance of the analyzer. The accepted range is typically about $30^\circ$ while electrons are emitted over 180$^\circ$ from a flat sample surface. It is possible to acquire photoelectrons throughout the full angular range by rotating the sample relative to the analyzer and performing multiple measurements in different geometries. This approach is not only time-consuming, but can also detract from the data quality: the sample rotation can modify the photoemission matrix elements, the polarization geometry, the beam position on the sample, and the absorbed energy density of the pump pulse in ultrafast experiments, all of which prevents a direct comparison of the obtained spectra.

These considerations are especially relevant for setups operating at comparatively low $\sim 10$~eV photon energies, which limit the momentum field of view intrinsically. Prominent examples and motivation for this work are time-resolved photoemission setups that use 6~eV probe photons which are generated by frequency upconversion of laser sources.\cite{Gauthier2020a} When 6~eV photons excite a metallic sample with 4~eV work function, electrons of up to 2~eV kinetic energy are emitted over a momentum range of 1.45~\AA$^{-1}$. This is only slightly smaller than the $\sim 1.6$~\AA$^{-1}$ Brillouin zone size of commonly studied materials with lattice constants of $\sim 4$~{\AA}. However, a $30^\circ$ field of view of conventional analyzers covers only 0.37~\AA$^{-1}$ momentum range. Collecting photoelectrons throughout the full angular range at once is therefore the most efficient way to observe the largest possible momentum range while keeping the benefits of low photon energies such as large photoemission cross-sections, least amount of space charge and increased bulk sensitivity for photon energies below $\sim10$~eV.\cite{M.P.Seah1979,Koralek07,Vishik2010,Tamai2019}

An electric field between the sample and the detector can be used to focus the photoelectron trajectories into the detector aperture and collect the complete angular range in a single, fixed geometry, see Fig.~\ref{fig1}. Such an electric field can be generated by applying a bias voltage to the sample or by applying an extractor voltage to the entrance aperture of the analyzer, creating a potential difference between the sample and the detector. The latter approach is employed in photoelectron microscopy systems\cite{Buseck1988} and more recently in momentum microscopes.\cite{Kromker2008,Tusche2015,Schonhense2015}

In contrast, the concept to bias the sample has rarely been used to measure a larger momentum range in systems with conventional hemispherical analyzers. However, sample bias has been applied for other purposes. For example, a sample bias was used in two-photon photoemission experiments in order to avoid complications at low kinetic energies,\cite{Rohleder2005,Muntwiler2007,Hengsberger2008} and it is useful to determine the work function of materials.\cite{Pfau2020,Ishida2020} The scarce use of this concept to expand the momentum field of view of conventional hemispherical analyzers might be explained by the challenge of establishing a well-defined and robust relation between the electron momenta in the solid and the measured angles that is applicable in the presence of an electric field. In particular, the electric field must be well-known to describe the photoelectron trajectories. Accordingly, recent implementations feature an electrically grounded mesh near the sample to minimize distortions of the electric field, however putting constraints on the incidence of the light source.\cite{Ichihashi2018,Yamane2019}

In this work, we describe an implementation of sample bias for a setup with a conventional hemispherical analyzer as widely used in the photoemission community. We demonstrate access to the largest emitted momenta in a single experimental geometry using a 6~eV laser light source. Our design, presented in section~\ref{secExpDet}, is  an upgrade that minimizes modifications of existing systems. Furthermore, in section~\ref{secModel} we establish an analytical, parameter-free model to adapt the angle-to-momentum conversion (Eq.~\ref{eq1}) to the effect of  a uniform electric field. Finally, in section~\ref{secExpRes} we demonstrate experimentally the validity of both the technique and the model using a representative dataset of the charge density wave (CDW) material TbTe$_3$. Our parameter-free model makes sample biasing  generally applicable to increase the momentum field of view while keeping a quantitative angle-to-momentum relation.

%%%%%%%%%%%%%%
%%% MODEL  %%%
%%%%%%%%%%%%%%
\section{Model}
\label{secModel}

\begin{figure*}[htb!]
\begin{center}
\includegraphics[scale=0.95]{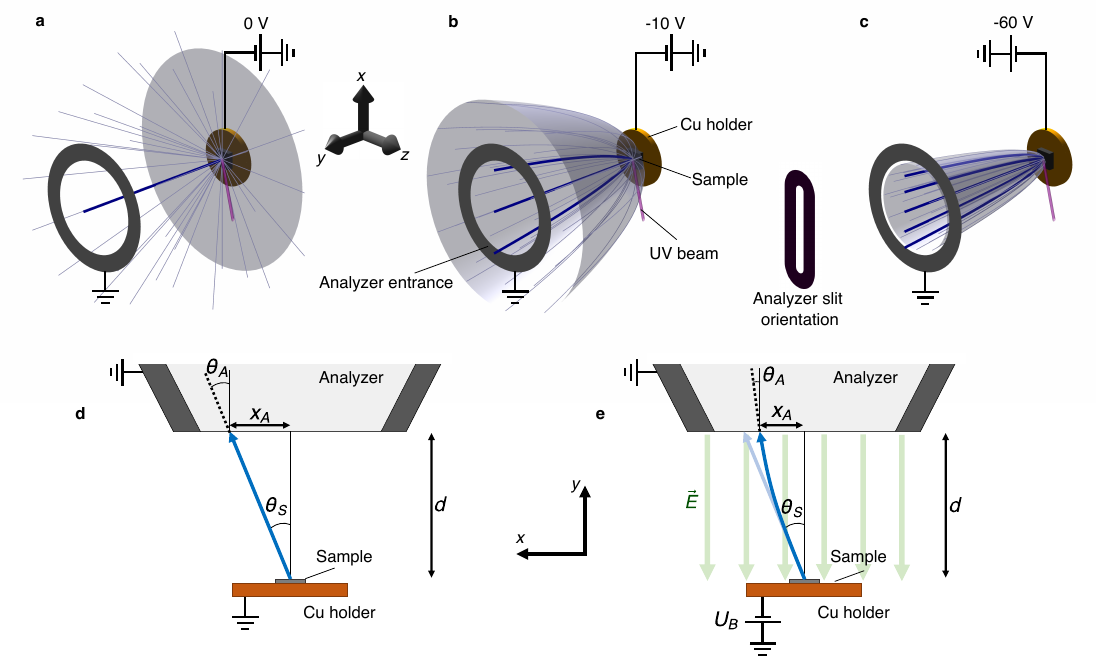}
\caption{(a) Illustration of the photoelectron trajectories emitted from the surface over a $2\pi$ solid angle. Only a small fraction of the electrons (indicated by thicker lines) pass through the circular aperture at the analyzer entrance (shown by a gray ring) as well as the vertical analyzer slit oriented along $x$ and are detected. (b)-(c) The photoelectron trajectories, considering the same initial conditions as (a), are accelerated towards the analyzer entrance when a negative bias voltage is applied. The emission cone is effectively focused towards the analyzer entrance. (d)-(e) Two-dimensional view of the photoelectron trajectories with identification of the relevant variables. Without bias voltage as in (d), the electron trajectory is straight and the angle of emission at the sample is the same at the analyzer entrance. With a bias voltage as in (e), the electron trajectory is curved by the electric field between the sample and the analyzer entrance. The emission angle and the angle at the analyzer entrance are then different. Note that a difference between the sample and analyzer work functions creates an additional electric field, which is included in our model by considering an effective bias voltage $U_B^*$. For the simplicity of the illustration, the work functions of the sample and the analyzer are assumed to be identical on this figure ($U_B^*=U_B$).
}
\label{fig1}
\end{center}
\end{figure*}
The angular range measured by a hemispherical analyzer is limited by the fixed aperture at its entrance. By applying a negative bias voltage between the sample and the electrically-grounded analyzer, an electric field is generated in the photoelectron flight path. This accelerates the electrons and bends their trajectories toward the analyzer entrance, therefore allowing for the detection of electrons with a larger photoemission angle, as schematized in Fig.~\ref{fig1}a-c. We are interested to establish a relation between the quantities measured by the detector and the ones at the sample surface in such a configuration. In a standard ARPES experiment without electric field, the angle $\theta_D$ and the kinetic energy $E_K^D$ recorded by the detector correspond directly to the emission angle $\theta_S$ and kinetic energy $E_K^S$ of the photoelectrons at the sample surface. In the presence of an electric field between the sample and the analyzer, this relation is modified. In the following, we present a simple model that establishes this modified relation and the angle-to-momentum conversion equation, similar to Eq.~\ref{eq1}, that applies in the presence of a uniform electric field. 

Throughout this work, we distinguish the quantities at the sample surface, at the analyzer entrance (before the electron lenses) and recorded by the detector with indices $S$, $A$ and $D$, respectively. We assume that the kinetic energy at the analyzer entrance and the one recorded by the detector are identical, i.e. $E_K^D=E_K^A$. In contrast, the angles $\theta_A$ and $\theta_D$ are not necessarily the same, as discussed later. 

First, we establish the relation between the kinetic energy at the sample surface $E_K^S$ and the one measured by the detector $E_K^D$. Those kinetic energies are given by
\begin{equation}
E_K^S=h\nu - \Phi_S - E_B
\end{equation}
\begin{equation}
E_K^D=h\nu - \Phi_A - E_B - eU_B,
\label{eq3}
\end{equation}
where $h\nu$ is the photon energy, $E_B$ is the electron binding energy in the solid, $U_B$ is the applied bias voltage and, $\Phi_S$ and $\Phi_A$ are the sample and analyzer work functions, respectively. This leads directly to the relation
\begin{equation}
E_K^S =E_K^D+ eU_B^*
\label{eqEK}
\end{equation}
where, for convenience, we have defined an effective bias voltage 
\begin{equation}
U_B^*=U_B+ \left(\Phi_A- \Phi_S\right)/e
\label{ubstar}
\end{equation}
which includes the difference between the sample and analyzer work functions. It is interesting to note that this work function difference generates an electric field even in the absence of bias voltage and should be generally considered, particularly at low photon energies and large differences of sample and analyzer work functions. 

With the energy relation established, we now turn to the angular relation. We approximate the electric field from the sample surface to the analyzer entrance by the uniform field generated in a parallel plate capacitor. This is applicable if the sample surface is normal to the electron lens axis, which is assumed throughout this work. With such a uniform electric field, the photoelectron momentum parallel to the sample surface is unchanged by the field, and consequently $k_{\parallel}^S = k_{\parallel}^A$. The parallel momentum at any position can always be expressed in terms of the kinetic energy and angle at that specific position. The angle-to-momentum conversion is then given in terms of quantities at the analyzer entrance by
\begin{equation}
k_{\parallel}^S = k_{\parallel}^A 
= \frac{1}{\hbar} \sqrt{2m E_{\text{K}}^A} \sin \theta_A.
\label{basic1}
\end{equation}
While the value of $E_K^A$ is identical to $E_K^D$, the relation between $\theta_A$ and $\theta_D$ remains to be established for this equation to be applicable.

Figure~\ref{fig1}d presents the configuration of a measurement in the absence of bias voltage, where $d$ 
is the distance between the sample surface and the analyzer entrance. This distance $d$ remains fixed to the analyzer focus distance throughout this work.
%represents the analyzer focus distance. 
In this configuration without bias voltage, the electron flight path is straight and the emission angle at the sample surface $\theta_S$ is equal to the angle $\theta_A$ measured at the analyzer entrance. By the design of the hemispherical analyzer, the angle $\theta_D$ measured by the detector is equal to $\theta_A$ in this standard configuration. We also define $x_A$ as the transverse position of the electron at the analyzer relative to normal emission. In Figure~\ref{fig1}d, the values $\theta_A$ and $x_A$ provide redundant information as they are related by
\begin{equation}
x_A=d\tan \theta_A.
\label{eq2}
\end{equation}
When the electrons are accelerated by an electric field, the photoelectron trajectory is bent as shown in Fig.~\ref{fig1}e. The values of $\theta_A$ and $x_A$ are both reduced in comparison to the field-free configuration. Eq.~\ref{eq2} is not valid anymore and the relation of $\theta_D$, $\theta_A$ and $x_A$ is not directly obvious in this case.  We can generally assume that $\theta_D = f(\theta_A,x_A)$. The unknown function $f$ represents the complex effect of the electrostatic lens before the hemispherical analyzer.\cite{Martensson1994,Wannberg2009} Note that our model only evaluates analytically the electron trajectories before the electrostatic lens and does not attempt to model the trajectories within the lens. To describe the effect of the electrostatic lens, we instead limit ourselves to two simple limits to approximate the function $f$ and verify their validity by comparing them to experimental results in section~\ref{secExpRes}. 
For both cases, we constrain $f$ such that it always correctly describes the zero-bias limit, i.e. $f(\theta_A,x_A)=\theta_A=\arctan(x_A/d)$.

\subsection*{Case I: Angular limit}
In the first case, we assume that only $\theta_A$ is important for the imaging process of the lens system. In this angular limit, we simply have 
\begin{equation}
\theta_D = f(\theta_A)=\theta_A. 
\end{equation}
It is then straight-forward to obtain an angle-to-momentum conversion equation from Eq.~\ref{basic1}: 
\begin{equation}
k_{\parallel}^S   
%= \frac{1}{\hbar} \sqrt{2m E_{\text{K}}^D} \sin \theta_A 
= \frac{1}{\hbar} \sqrt{2m E_{\text{K}}^D} \sin \theta_D.
\label{eqkconvCaseI}
\end{equation}
This is simply the conventional conversion equation based on the quantities measured by the detector. Our results in section \ref{secExpRes} indicate that this conversion is incorrect. 

For later comparison with the case II, we rewrite the angle-to-momentum conversion in terms of the kinetic energy at the sample surface:
\begin{equation}
 k_\parallel^S= \frac{1}{\hbar} {\sqrt{2mE_K^S}} F \sin\theta_D.
\end{equation}
Here, we defined the momentum-scaling factor $F$ as
\begin{equation}
F=\sqrt{1+2\alpha}
\label{FfactorCaseI}
\end{equation}
with the kinetic parameter $\alpha$ defined in Eq.~\ref{eqAlpha}.

\subsection*{Case II: Position limit}
In the second limit, we assume that only the value of $x_A$ is important for the imaging process of the lens system. 
In this limit, the function $f$ is given by
\begin{equation}
\theta_D = f(x_A)=\arctan(x_A/d).
\label{Eqatan}
\end{equation}
The value of $x_A$ can be evaluated from basic electron kinematics: 
%as shown below. 
%The electron position $\vec{x}(t)$ is given by:
%\begin{equation}
%\vec{x}(t)=\vec{v} t + \frac{1}{2} \vec{a} t^2
%\end{equation}
%where $t$ is the time, $\vec{v}=(v_x,v_y)$ is the initial velocity and $\vec{a}=(0,a)$ is the acceleration with the $xy$ axes defined in Fig.~\ref{fig1}d-e. At a time $t_A$, the electron will reach the analyzer at a position $\vec{x}(t_A)=(x_A,d)$. The value of $x_A$ is therefore given by:
\begin{equation}
x_A=\frac{v_x}{a}\left({ \sqrt{v_y^2+2ad}} -v_y \right)
\end{equation}
where $\vec{v}=(v_x,v_y)$ is the initial velocity and $\vec{a}=(0,a)$ is the acceleration with the $xy$ axes defined in Fig.~\ref{fig1}d-e with the analyzer slit oriented along the $x$-axis. 

Considering Eq.~\ref{Eqatan} and $v_x^2+v_y^2=2E_K^S/m$, we find
\begin{equation}
\tan \theta_D=\frac{v_x}{ad}\left({\sqrt{\frac{2E_K^S}{m}-v_x^2+2ad} -\sqrt{\frac{2E_K^S}{m}-v_x^2} }\right).
\label{eqBigOne}
\end{equation}
Solving this equation for $v_x$, keeping only the solution which is non-zero in the limit $a \rightarrow 0$
%valid in the limit of $a=0$ 
and rewriting it in terms of the parallel momentum with $mv_x=\hbar k_{\parallel}^S$, we obtain
\begin{equation}
 k_\parallel^S=
 \frac{1}{\hbar} {\sqrt{2mE_K^S}} F \sin\theta_D
\label{eqKconv}
\end{equation}
where the momentum-scaling factor $F$ is defined as
\begin{equation}
F=\sqrt{\frac{\alpha+1+ \sqrt{2\alpha +1-\alpha^2\tan^2\theta_D }  }{2} }
\label{eqConv}
\end{equation}
with the kinetic parameter $\alpha$ being
\begin{equation}
\alpha = \frac{mad}{2E_K^S}.
\end{equation}

The acceleration $a$ is caused by the electric field generated by the effective bias voltage $U_B^*$, combining the applied bias voltage and the difference between the sample and analyzer work functions (see Eq. \ref{ubstar}). Considering the acceleration $a=-eU_B^*/md$, the kinetic parameter $\alpha$ is rewritten as
\begin{equation}
\alpha = \frac{-eU_B^*}{2E_K^S}.
\label{eqAlpha}
\end{equation}

The angle-to-momentum conversion for the position limit (case II) is therefore formed by the set of equations~\ref{eqEK}, \ref{ubstar}, \ref{eqKconv}, \ref{eqConv} and \ref{eqAlpha}. We demonstrate in section~\ref{secExpRes} that this conversion can be successfully applied to our experimental results. Note that these equations are only physically relevant for positive values of the velocity $v_y$. In the limit $v_y \rightarrow 0$, the kinetic energy at the sample is only given by the parallel momentum $k_\parallel^S$ and defines a low-energy cutoff (LEC) in the photoemission spectrum at $E_{K,LEC}^S=\hbar^2 (k_\parallel^S)^2/2m$. This LEC recorded by the detector can be expressed as
\begin{equation}
%\theta_D^{LEC} = \pm \arctan \left( 2\sqrt{\frac{E_{K,LEC}^S}{-eU_B^*}} \right)
%E_{K,LEC}^S = -eU_B^* \left( \frac{\tan \theta_D}{2} \right)^2 
E_{K,LEC}^D =-eU_B^*\left[1 +\left( \frac{\tan \theta_D^{LEC}}{2} \right)^2 \right]
\label{eqLEC}
\end{equation}
by taking $2E_K^S/m=v_x^2$ in Eq.~\ref{eqBigOne}.

\begin{figure}[htb!]
\begin{center}
\vspace{0.5cm}
\includegraphics[scale=1]{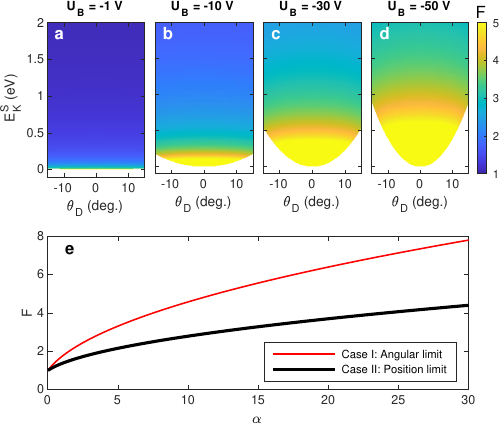}
\caption{(a)-(d) Momentum-scaling factor $F$ in the position limit (Eq.~\ref{eqConv}) for relevant ranges of detector angle $\theta_D$ and kinetic energy $E_K^S$ at various bias voltage with $\Phi_A=\Phi_S$. The angular dependence is weak while the energy dependence is pronounced near the low-energy cutoff (LEC). The factor $F$ is undefined below the LEC, as no photoelectrons can physically have those parameters. (e) Comparison of the momentum-scaling factor $F$ at $\theta_D=0$ for the angular (case I) and the position (case II) limits with Eqs.~\ref{FfactorCaseI} and \ref{eqConv}, respectively. 
%hv=6; %eV
%WFa=4.15;
%WFs=WFa;
}
\label{fig3}
\end{center}
\end{figure}

In Fig.~\ref{fig3}, we illustrate the momentum-scaling factor $F$ (Eq.~\ref{eqConv}) for different values of bias voltage. The momentum-scaling factor $F$ is an indicator of the increased momentum range relative to a measurement without bias voltage. For example, the momentum range measured for $E_K^S=1$~eV is about four times larger at $U_B=-50$~V (Fig.~\ref{fig3}d) in comparison to 0~V. Overall, we see that for a given bias the factor $F$ is mostly dependent on $E_K^S$ and changes rapidly near the LEC. It has only a weak dependence on $\theta_D$. The size of $F$ becomes more important for larger bias voltages, representing a larger field of view in momentum space.

To compare the angle-to-momentum conversion from cases I and II, we trace their respective momentum-scaling factor (Eqs.~\ref{FfactorCaseI} and \ref{eqConv}) in Fig.~\ref{fig3}e at $\theta_D=0$. It shows that the measured momentum range, in identical conditions, would be at least 50\% larger for case I than for case II when $\alpha>5$. We show in section~\ref{secExpRes} that the position limit (case II) is consistent with our experimental results while the angular limit (case I) overestimates the momentum values.

\subsection*{Scaling of photoemission intensities}

Experimentally, the measured quantity is the photoemission intensity as function of angle $N(E_K^D,\theta_D)$ while we are interested in the intensity as function of momentum $N(E_K^S,k_\parallel^S)$. The coordinate transformation from $\theta_D$ to $k_\parallel^S$ derived above is accompanied by a transformation of the differential line element $d\theta_D \rightarrow dk_\parallel^S$ that causes a scaling of the intensity:
\begin{equation}
\frac{N(E_K^S,k_\parallel^S)}{\Delta k_\parallel^S}
=
\frac{N(E_K^D,\theta_D)}{\Delta \theta_D}
\left|\frac{\partial k_\parallel^S}{\partial \theta_D} \right|^{-1}
\label{eqIntensityScaling}
\end{equation}
where $\Delta k_\parallel^S$ and $\Delta \theta_D$ are the bin sizes. The partial derivative from Eqs.~\ref{eqKconv}-\ref{eqConv} is given by
\begin{equation}
\frac{\partial k_\parallel^S}{\partial \theta_D} = 
\frac{\sqrt{2mE_K^S}}{\hbar}
\left[ F \cos \theta_D + \sin \theta_D \frac{\partial F}{\partial \theta_D}
\right]
\end{equation}
with 
\begin{equation}
\frac{\partial F}{\partial \theta_D} =
\frac{-\alpha^2 \tan \theta_D \sec^2 \theta_D}{4F \sqrt{2 \alpha +1 -\alpha^2 \tan^2 \theta_D}}.
\end{equation}

In general, the largest intensity correction occurs near the LEC. We note that this intensity scaling resulting from the coordinate transformation is a generic feature, present even without a bias voltage, and should generally be considered in any standard ARPES experiment, particularly when studying several eV wide spectra.

Without a bias voltage or difference of analyzer and sample work functions ($U_B^*=0$), the partial derivative is simply given by 
\begin{equation}
\frac{\partial k_\parallel^S}{\partial \theta_D} = 
\frac{1}{\hbar} {\sqrt{2mE_K^S}}
 \cos \theta_D.
 \end{equation}

%%%%%%%%%%%%%%
%%% EXP  %%%
%%%%%%%%%%%%%%
\section{EXPERIMENTAL IMPLEMENTATION}
\label{secExpDet}

\begin{figure}[htb!]
\begin{center}
\includegraphics[width=\columnwidth]{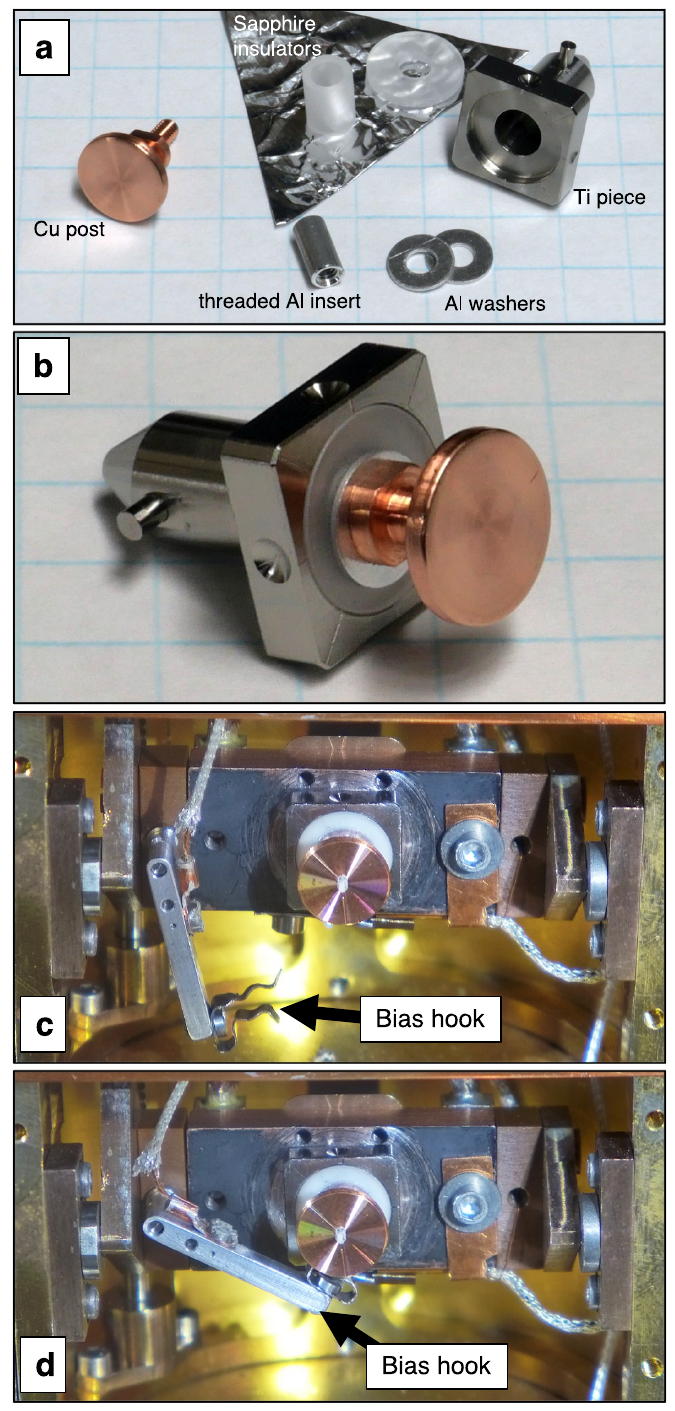}
\caption{Photographs of the implementation of sample bias. (a)~Components of the electrically insulated mount, including sapphire pieces to electrically decouple the titanium piece from the threaded aluminum insert. (b)~Assembled sample mount with the copper post, onto which samples are fixed. (c)~Sample holder on the manipulator. The bias hook is removed to allow sample transfer. (d) The bias hook is attached to apply bias voltage.}
\label{figImp}
\end{center}
\end{figure}

To realize the scheme discussed above, the experimental setup must allow the application of a voltage on the sample, which must be insulated from the electrically-grounded manipulator. The insulator must support operation at cryogenic temperatures. The geometry of the design must also generate a region with uniform electric field to avoid any distortions of the photoemission spectra. 

We implemented these requirements in our experimental setup in the following way. The bias voltage is generated from a DC Voltage Source DC205 from Stanford Research Systems and applied to the sample through a wire running from the top of the manipulator to the sample stage. The Kapton-insulated wire is electrically shielded by a grounded silver wire braid to avoid charging issues due to stray electrons. On the sample stage, this wire is connected to a hook made from titanium sheet that can be attached to the copper post onto which the sample is fixed~(Fig.~\ref{figImp}c-d). 

The OFHC copper post is screwed into a threaded 6061 aluminum insert, which is insulated from the titanium mounting piece by a sapphire cylinder and washer (Insaco Inc), (Fig.~\ref{figImp}a-b). Sapphire insulators are chosen to optimize cryogenic performance. The combination of titanium and sapphire is favorable due to similar thermal expansion coefficients.\cite{Marquardt02} The threaded aluminum insert and titanium piece are individually machined to match the sapphire inserts with tight $<10$~$\mu$m tolerances. This assembly is fastened by application of a thin film of Loctite Stycast 1266 that has been outgassed in high vacuum before curing to achieve compatibility with ultra-high vacuum. The assembly is cured in a jig which pulls on the threaded aluminum insert while pushing down on the sapphire washer and cylinder to ensure tight fits with optimal thermal contact. Aluminum washers are used to set the sample post orientation with respect to the titanium piece.

In order to form a parallel plate capacitor geometry, the top surface of the copper post is designed as a disk with a 1~cm diameter, the largest diameter  possible to allow sample transfer in our experimental setup. The analyzer entrance is electrically grounded and acts as the second plate of the capacitor. The measurements are taken for a configuration near normal emission to maintain the parallel plate geometry and keep a uniform electric field. 

In these experiments a Spectra Physics Ti:Sapphire oscillator operating at a repetition rate of 80~MHz generates 1.5~eV photons that are frequency quadrupled through two stages of second harmonic generation in $\beta$-BaB$_2$O$_4$ non-linear crystals to provide a 6~eV photon source.\cite{Kato1986,Gauthier2020a} The photoelectrons are detected by a Scienta R4000-WAL hemispherical analyzer, which has a work function $\Phi_A=4.15$~eV (determined using Eq.~\ref{eq3} with the measured value of $E_K^D$ at the Fermi level). The analyzer slit width was fixed to 0.1~mm throughout the measurements. Measurements were performed using a pass energy of 10~eV in dither mode. The distance $d$ between the sample and the analyzer entrance was fixed to the instrument-defined focus distance throughout the measurements.

To demonstrate the effects of sample bias and the validity of our model, we performed measurements on the compound TbTe$_3$, a member of the well studied rare-earth tritelluride family featuring CDW order.\cite{Brouet2008,Ru2008} TbTe$_3$ single crystals were grown using a Te self-flux technique,\cite{Ru2006} which ensures purity of the melt, and produces large crystals with a high degree of structural order. Elements in the molar ratio Tb:Te~$=0.03/0.97$ were put into alumina crucibles and vacuum sealed in quartz tubes. The mixture was heated to 900$^\circ$C over the course of 12h and kept at that temperature for a further 10h. It was then slowly cooled to 650$^\circ$C over a period of 100h. The remaining melt was decanted in a centrifuge. The resulting copper-colored crystals are malleable plates with dimensions of up to $5\times5\times0.4$~mm, and oriented with the long $b$ axis perpendicular to the plane of the crystal plates. The nearly equal in-plane axes $a$ and $c$ are parallel to the crystal growth edges, but must be distinguished using for instance x-ray diffraction. The material is air sensitive, and crystals were stored in an oxygen and moisture-free environment. 

TbTe$_3$ was cleaved in situ at a base pressure of $1\times10^{-10}$~Torr and measured at a constant temperature of 84~K. Measurements were performed for $U_B$ ranging from 0~V to $-60$~V. The 6~eV flux was kept constant and its weak time-dependent drift was corrected for in the analysis. The measurement position was optimized at $U_B=-60$~V for normal emission, i.e. for the largest energy difference between the Fermi level ($E_\mathrm{F}$) and the minimum of the LEC, as justified in the end of section \ref{secExpRes}. The position remained fixed for measurements at all bias voltages. The sample work function was determined to be 5.14~eV. 

\section{Experimental results}
\label{secExpRes}

\begin{figure}[htb!]
\begin{center}
\includegraphics[scale=0.97]{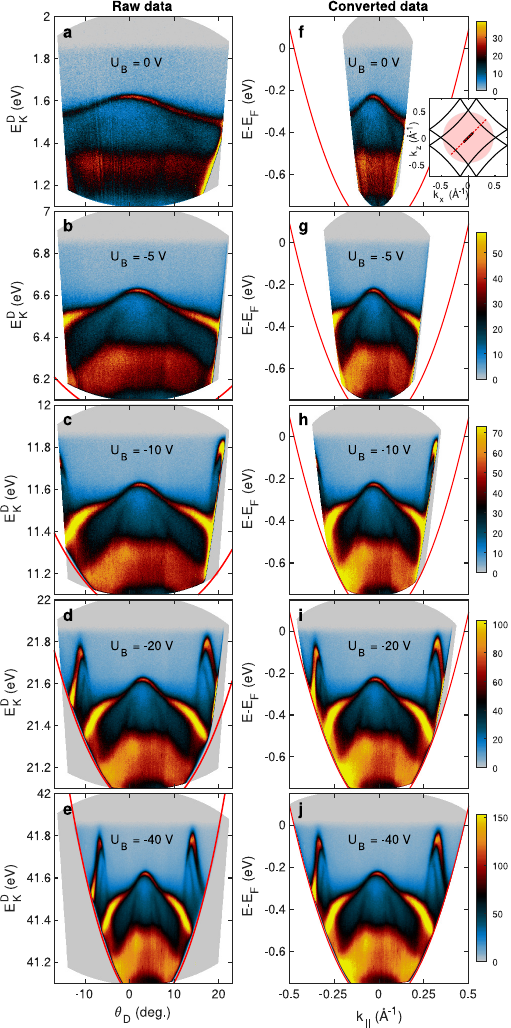}
\caption{(a)-(e) Raw photoemission spectra for $U_B=0$~V to $-40$~V. The red line indicates the low energy cutoff (LEC) defined by Eq.~\ref{eqLEC}. (f)-(j) Corresponding photoemission spectra after angle-to-momentum conversion (Eqs.~\ref{eqKconv}-\ref{eqConv}) and intensity scaling (Eq.~\ref{eqIntensityScaling}). The red line indicates the LEC, below which the angle-to-momentum conversion is not defined. Inset in (f): Schematic Fermi surface of TbTe$_3$. The red circle indicate the momentum at $E_\mathrm{F}$ accessible with 6~eV photons. The solid black and dashed red lines indicate the measured cuts at 0~V and $-40$~V, respectively.}
\label{fig4}
\end{center}
\end{figure}

Measurements were performed in the $k_x-k_z$ plane with the analyzer slit along the diagonal of the Brillouin zone of TbTe$_3$, as illustrated in the inset of Fig.~\ref{fig4}f. Here, $k_x$ and $k_z$ indicate the reciprocal axes of the TbTe$_3$ crystal axes $a$ and $c$, respectively. The red circle illustrates the physically accessible momentum at $E_\mathrm{F}$, limited by the LEC based on the sample work function and photon energy. The solid red line represents the momentum range observable through the $30^\circ$-wide aperture at $U_B=0$~V while the dashed line indicates the same at $U_B=-40$~V according to our model. 

The measured raw spectra from $U_B=0$ to $-40$~V are presented in Fig.~\ref{fig4}a-e. The central hole-like band located at $\approx-0.25$~eV below $E_F$ is compressed on the angular axis as a bias voltage is applied. Furthermore, for $U_B<-10$~V new steeply dispersing bands appear on the edge of the spectra. These results clearly demonstrate that photoelectrons with larger parallel momenta are focused into the analyzer by the electric field. The theoretically expected LEC from our model (case II, Eq. \ref{eqLEC}) is shown as a red line in Fig.~\ref{fig4}a-e and is in remarkable agreement with the experimental cutoff. The effective normal-emission angle $\theta_0$ of our data is determined from the position of the LEC minimum. 

The data, converted in momentum space using Eqs.~\ref{eqKconv}-\ref{eqConv} and its intensity scaled according to Eq.~\ref{eqIntensityScaling}, are presented in Fig.~\ref{fig4}f-j. The wider field of view in momentum space with larger $|U_B|$ is obvious. More importantly, the angle-to-momentum conversion, without any free parameter, results in spectral features that are mostly independent of $U_B$. This supports the validity and the applicability of the model. In the following, we present an analysis of various spectral features to illustrate the accuracy of the conversion as well as the limitations of the technique. 

\begin{figure}[htb!]
\begin{center}
\includegraphics[scale=0.97]{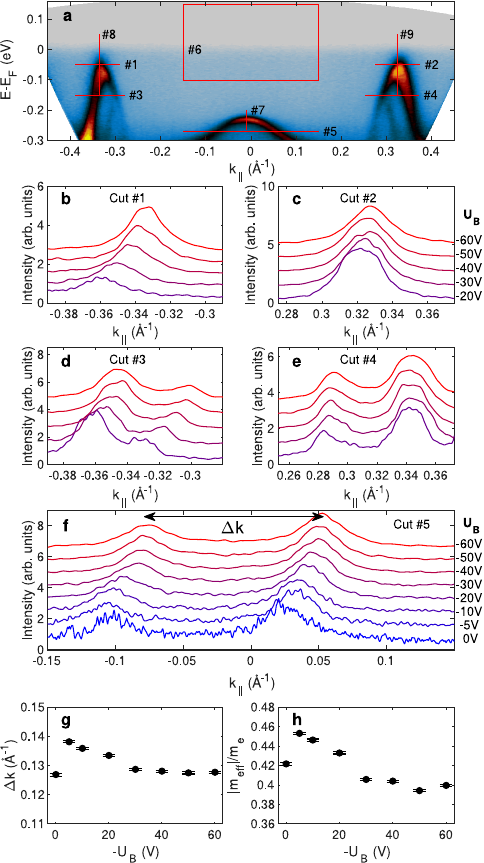}
\caption{
(a) Photoemission spectrum at $U_B=-60$~V with various cuts (shown in Figs.~\ref{fig5}, \ref{fig6} and \ref{fig7}) indicated in red. (b)-(e) MDCs of the steeply dispersing bands near $k_\parallel \approx 0.34$~\AA$^{-1}$ as a function of $U_B$. (f) MDCs of the central hole band as a function of $U_B$. (g) Peak separation $\Delta k$ of the central hole band from MDCs in (f) as a function of $U_B$. (h) Effective mass of the central hole band as a function of $U_B$.}
\label{fig5}
\end{center}
\end{figure}

In TbTe$_3$, the CDW order opens a gap on large portions of the Fermi surface, with the largest gap occurring for $k_x=0$ and decreasing magnitude with increasing $|k_x|$.\cite{Brouet2008,Schmitt2008} Along the cut shown in the inset of Fig.~\ref{fig4}f, the gap is still finite and should appear at $|k_\parallel^S|\approx0.34$~\AA$^{-1}$. These expectations are in good agreement with our experimental results. In particular, a band gap, together with its band backfolding, is clearly observed for $U_B=-40$~V at the expected momentum. In Fig.~\ref{fig5}, we present momentum distribution curves (MDCs) of significant features as a function of $U_B$, with various cuts identified in Fig.~\ref{fig5}a. Cuts 1 and 2 (Fig.~\ref{fig5}b-c) show MDCs at the top of the gapped band for positive and negative momenta while cuts 3 and 4 (Fig.~\ref{fig5}d-e) present MDCs illustrating the main band and its associated back-folded band. These features, observed from -20~V to -60~V, appear near the expected momenta but are not completely independent of $U_B$. We attribute this weak $U_B$-dependence to experimental limitations, as discussed in details at the end of this section. In order to characterize the bias voltage dependence down to $U_B=0$~V, cut 5 presents MDCs of the hole-like band centered at $k_\parallel=0$ (Fig.~\ref{fig5}f). The separation $\Delta k$ between both peaks is nearly constant with $U_B$~(Fig.~\ref{fig5}g). There is however a clear $U_B$-dependent peak position shift that we again attribute to experimental limitations. To further illustrate the validity of the angle-to-momentum conversion model, we also present the mass of the hole band in Fig.~\ref{fig5}h. To determine the hole mass, the band dispersion was extracted by fitting energy distribution curves (EDCs) over the range of $\Delta k$. 
The effective hole mass does not deviate from the zero-bias value by more than 10\% at all $U_B$.
%The effective hole mass is mostly independent of $U_B$, with some deviations for small $|U_B|$. 
This result is remarkable considering that our simple angle-to-momentum conversion model has no adjustable parameter. 

We note that the angular limit (case I) leads to a momentum-scaling factor at $U_B=-60$~V about twice as large as the one used here for the position limit (case II). It is therefore clear that the angular limit would lead to strongly $U_B$-dependent spectral features. Furthermore, it would result in bands at momentum values larger than physically allowed from the sample work function and photon energy of the experiment. We can therefore conclude that the angular limit (case I) is not valid in the experimental conditions considered in this work. 

We note that the increase in momentum range by a factor $F$ affects both the in-plane momentum parallel and transverse to the analyzer slit. This results in integrating a range of transverse momentum $F$ times larger than without bias voltage, relaxing the transverse momentum resolution but also increasing the count rate by a factor $F$. This increase in count rate at a constant 6~eV flux is directly evident from the color scale on Fig.~\ref{fig4} and signal to noise ratio of the cuts in Fig.~\ref{fig5}. Depending on the specific scientific question, this effective increase in analyzer transmission can be an additional important benefit. 

\begin{figure}[htb!]
\begin{center}
\includegraphics[scale=1]{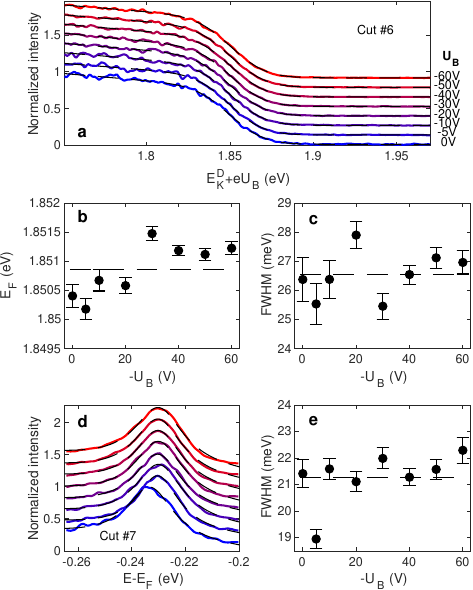}
\caption{(a) EDCs illustrating the Fermi edge caused by secondary electrons at the center of the Brillouin zone for various bias voltages. The data were fitted with a Fermi-Dirac distribution convoluted with a Gaussian function. The determined $E_\mathrm{F}$ and Gaussian FWHM are presented in (b) and (c), respectively. (d) EDCs of the central hole band as a function of $U_B$. The data were fitted to a Lorentzian function with a linear background. The determined Lorentzian FWHM is presented in (e). }
\label{fig6}
\end{center}
\end{figure}

We further characterize how the bias voltage affects spectral features on the energy axis. We first evaluate the intensity near $E_\mathrm{F}$. Without any bands crossing $E_\mathrm{F}$ for all $U_B$, in Fig.~\ref{fig6}a we instead analyze the Fermi edge caused by electrons scattered in the final state. This region is marked in the center of the Brillouin zone as box 6 in Fig.~\ref{fig5}a. The Fermi edge was fitted to a Fermi-Dirac function convoluted with a Gaussian resolution function. The resulting position of $E_F$ and the full-width-at-half-maximum (FWHM) of the resolution function are presented in Fig.~\ref{fig6}b and c, respectively. Also, the EDCs of the central hole-like band were fitted to a Lorentzian function with a linear background and the resulting FWHM is presented in Fig.~\ref{fig6}e. Overall, we observe no significant energy shift or broadening due to the bias voltage, therefore demonstrating that the technique does not cause significant artefacts along the energy-axis.

Finally, we discuss experimental limitations of the bias voltage technique. The model is based on the assumption that a uniform electric field exists from the sample surface until the analyzer. However, distortions of the electric field will occur when the configuration deviates from a parallel plate capacitor. In particular, the bias application apparatus can cause stray field and great care must be taken when designing the biasing mechanism to shield electric fields to the extent possible. Furthermore, difference in work functions of the sample and of the holder can also create inhomogeneous fields near the sample surface.\cite{Fero2014} Note that this effect is unrelated to the sample bias and is an issue for any ARPES measurements of photoelectrons with low kinetic energy. In the case here, the sample work function of 5.14~eV is sufficiently different from the one of the copper sample holder ($\approx4.7$~eV) to cause sizable field distortions. In practice, one can try to compensate the distortions in the electric field by adjusting slightly the sample position and/or orientation. Routine use of sample bias therefore also requires a precise and accurate sample manipulator. 

\begin{figure}[htb!]
\begin{center}
\includegraphics[scale=1]{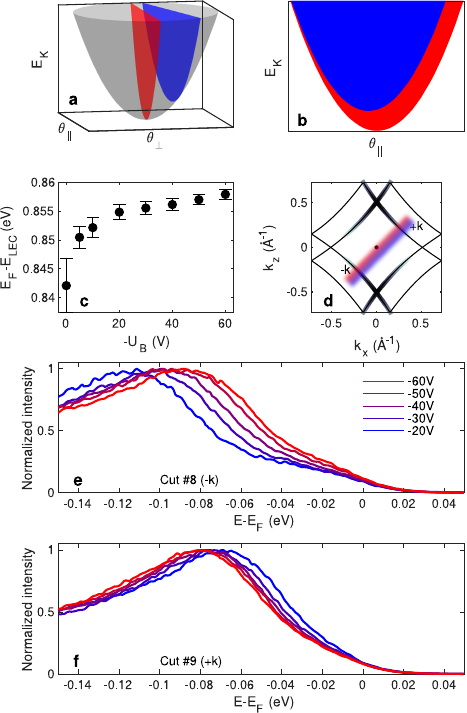}
\caption{(a) Schematic representation of the low energy cutoff (LEC) in gray for photoemission angles parallel  ($\theta_\parallel$) and perpendicular ($\theta_\perp$) to the analyzer slit. The red and blue areas illustrate measurements taken at normal emission and off-normal emission, respectively. The lowest possible kinetic energy occurs for normal emission (red), as shown more clearly by the projection on $\theta_\parallel$ in (b). (c) Experimental energy difference between $E_\mathrm{F}$ and the minimum of LEC. For perfectly normal emission, this quantity is equal to $h\nu - \Phi_S$. (d) Schematic of the Fermi surface where the line thickness represents the size of the CDW gap which is maximal at $k_x=0$. A cut going directly through the center of the Brillouin zone (red region) exhibits equal gaps at $+k$ and $-k$, while a cut slightly displaced away from it (blue region) exhibits different gaps at $+k$ and $-k$. (e)-(f) EDCs illustrating an increase (decrease) of the gap at $-k$ ($+k$) for increasing $|U_B|$. This change indicate that the effective cut in momentum space changes as 
illustrated in (d) as a function of $U_B$.}
\label{fig7}
\end{center}
\end{figure}

Our model can only be applied for normal emission measurements as it relies on a parallel plate capacitor geometry. We therefore optimize for normal emission using the LEC. Specifically, we optimize for the lowest measurable kinetic energy defined by electrons without any in-plane momentum, as schematized by the red parabola in Fig.~\ref{fig7}a,b. In the presence of field distortions,  electrons acquire a finite in-plane momentum transverse to the analyzer slit and one measures effectively off-normal emission while sample and analyzer are oriented parallel. This is indicated by a  higher energy minimum of the LEC, as illustrated with the blue parabola in Fig.~\ref{fig7}a,b. One can therefore optimize the experimental geometry to minimize the LEC energy relative to $E_\mathrm{F}$ to compensate for field distortions. This ensures that the measured electrons at $k_\parallel=0$ are emitted normal to the sample surface.

In the measurements presented in this work, we optimized the LEC at a bias voltage $U_B=-60$~V and confirmed that the spectrum near $E_\mathrm{F}$ is symmetric around $k_\parallel=0$ after optimizing the geometry. This geometry was then retained for all bias values to provide a direct comparison. However, the field compensation changes with bias and the effective cut in momentum space can be modified. We indeed observe this effect, as illustrated in Fig.~\ref{fig7}.  In Fig.~\ref{fig7}c, we present the energy difference between $E_\mathrm{F}$ and the minimum of the LEC. This quantity is maximal for $U_B=-60$~V and decreases with decreasing $|U_B|$. This is a signature that the measured electrons acquire a finite momentum transverse to the analyzer slit with decreasing $|U_B|$. The effective cut in momentum space therefore changes from a cut going through the center of the Brillouin zone at $U_B=-60$~V to a displaced cut for smaller $|U_B|$. This effect explains the observed asymmetry of the CDW gap for bias voltages different from  $U_B=-60$~V (Fig.~\ref{fig7}e,f). Indeed, an asymmetry is expected for cuts in momentum space displaced from the center of the Brillouin zone. In Fig.~\ref{fig7}d, the thickness of the black lines represent the size of the CDW gap in TbTe$_3$. While it is identical for positive ($+k$) and negative ($-k$) momentum for a cut going through the center of the Brillouin zone (red region), it will be larger at $-k$ than at $+k$ for a cut away from it (blue region). The combined observation of the CDW gap asymmetry and of the change in the LEC minimum strongly supports our interpretation that the effective cut in momentum space is modified with $U_B$. This change in the cut also explains some variations of the spectral features observed with different bias voltage in Fig.~\ref{fig5}.

%%%%%%%%%%%%%%
%%% CONCLUSIONS %%%
%%%%%%%%%%%%%%
\section{Discussion and Conclusions}
\label{secDis}

Our experimental results show that the momentum field of view in ARPES experiments with hemispherical analyzers can be increased by applying a bias voltage to the sample. In addition, we derive an analytic, parameter-free expression for the conversion between measured angles and in-plane momenta that is applicable when the electric field is uniform. We confirm its validity with our experimental results on TbTe$_3$. As our model is parameter-free, it allows us to perform bias-measurements for a wide range of conditions without the need for any calibration. 

Furthermore, the model has important implications even for measurements without bias voltage. Specifically, if the difference in sample and analyzer work functions is comparable to the photoelectron kinetic energies, the induced electric field will have observable effects on the photoelectron trajectories and our model should be considered to obtain accurate momentum and intensity values. Such a regime is commonly encountered in ARPES measurements using 6~eV photon energy.

We like to point out another approach to establish an angle-to-momentum conversion for a comparable experimental configuration. Jauernik \textit{et al}. developed a heuristic model that was calibrated to the well-known dispersion of the image potential state in front of the (001) surface of copper.\cite{Jauernik2018} The empirical bias-scaling factor of their model can be estimated using our parameter-free model. We find good agreement to the experimentally determined value, further supporting the general validity of our model.\footnote{The authors in Ref.~\onlinecite{Jauernik2018} determined a bias-scaling factor $k\approx0.4$ in their empirical model based on their experimental results. Considering the experimental parameters $E_K^S=4.2$, $U_B=-10$~V and assuming $\Phi_A=\Phi_S$, our parameter-free model predicts a bias-scaling factor $k\approx0.42$, in good agreement with their findings.} 

While our model is limited to normal emission by definition, it could be generalized to off-normal emission by assuming that the electric field is generated by two infinite planes that intersect at a line.~\cite{Hengsberger2008} 

The main experimental limitation of our implementation of sample bias is the distortion of the photoemission spectra caused by deviations from a uniform electric field, as defined by a parallel plate capacitor geometry. The general design of an ARPES system can limit the validity of this approximation, as well as the specific characteristics of different samples. As we demonstrate, a symmetric bias field between the sample, its holder and the analyzer can be obtained by careful engineering of the experimental design. Other objects near the photoelectron trajectory, such as the capillary of a Helium lamp can cause significant field distortions\cite{Ichihashi2018} and a different design with its own angle-to-momentum conversion formalism might be necessary.\cite{Ichihashi2018,Yamane2019} We also note that samples with large flat surfaces are preferred to obtain a uniform electric field. The investigation of samples with rough surfaces is more challenging due to irregular fields near the surface. A photoemission spot size smaller than the characteristic length scale of the sample inhomogenities is beneficial. Field distortions can also be minimized by reducing the work function differences between the sample and its holder.\cite{Fero2014} As evidenced in Fig.~\ref{fig7}, the experimental configuration should be optimized for each specific bias voltage to compensate the field distortions that remain and obtain the most reliable results. Consequently, precise and stable positioning of the sample becomes more important as well.

A fundamental physical limitation of our approach remains the intrinsically limited momentum range at low photon energies. While it is possible to continue to compress the momentum range by increasing the bias voltage value, the maximum accessible momentum remains physically limited by the photon energy and the sample work function. For example, in our experiment the complete physically allowed momentum range at $E_\mathrm{F}$ is observed for $U_B=-40$~V and larger $|U_B|$ values do not provide additional information. 

ARPES experiments at $\sim 10$~eV photon energies profit most from the increased momentum field of view a sample bias provides. A fixed sample orientation avoids  issues caused by matrix elements, polarization effects, and beam walk on small samples, while retaining the advantages of high photoemission intensity, enhanced bulk sensitivity and mitigation of space charge effects at low photon energies. The application of a bias voltage is not limited to photoemission with 6\,eV photons used in this study. In this context we note the recent development of an ultrafast 11~eV laser for time-resolved studies\cite{Lee2020,Peli2020} and a quasi-continuous wave 11~eV laser for high resolution ARPES.\cite{He2016,Tamai2019}  

Time-resolved ARPES experiments relying on 6~eV probe photon energies particularly benefit from measuring the electron dynamics over a large part of the Brillouin zone in a fixed configuration. In addition to all of the benefits applying in equilibrium, a fixed geometry avoids changes in absorbed excitation density in pump-probe experiments.\cite{Yang2019} Furthermore, the notable increase in effective analyzer transmission due to a larger acceptance range of transverse momenta can be an important aspect when collecting statistics sufficient for high precision studies.

%In this context, it facilitates orbital tomography studies for photon energies in the tens of electronvolts.\cite{Yamane2019}

\begin{acknowledgments}
This work was supported by the Department of Energy, Office of Basic Energy Sciences, Division of Materials Science and Engineering.
N.G. acknowledges support from the Swiss National Science Foundation (fellowship no. P2EZP2 178542).
H.P. acknowledges support from the German Science Foundation (DFG) under reference PF 947/1-1 and from the Advance Light Source funded through U.S. Department of Energy, Office of Science, Office of Basic Energy Sciences, under Contract No. DE-AC02-05-CH11231. 
M.D.B. acknowledges partial support from the Swiss National Science Foundation under project number P2SKP2 184069, as well as from the Stanford Geballe Laboratory for Advanced Materials (GLAM) Postdoctoral Fellowship program.
\end{acknowledgments}

\section*{DATA AVAILABILITY}
The data that support the findings of this study are openly available in the Stanford
Digital Repository at \href{http://doi.org/10.25740/sk226xw9348}{http://doi.org/10.25740/sk226xw9348} (Ref. \cite{Gauthier2021}).

\section*{References}
\bibliography{refBMC.bib}

\end{document}